\documentclass[conference]{IEEEtran}
\ifCLASSINFOpdf
\else
\usepackage[dvips]{graphicx}
\fi
\usepackage[cmex10]{amsmath}
\usepackage{upgreek}
\usepackage{booktabs}
\usepackage{epsfig}
\usepackage{latexsym}
\usepackage{multirow}
\usepackage{stfloats}
\usepackage{epstopdf}
\usepackage{color}  
\usepackage{tabularx} 
\usepackage{amssymb}
\usepackage{enumerate}
\graphicspath{{./Figures/}}
\usepackage{color}
\usepackage{bbm}
\usepackage{bm}
\usepackage{cite}
\usepackage[tight,footnotesize]{subfigure}
\usepackage{balance}
\usepackage{mathrsfs}
\usepackage{verbatim}
\usepackage{dsfont}
\usepackage{verbatim}
\usepackage{tikz}
\usepackage{diagbox}
\usepackage{caption}
\allowdisplaybreaks[4]
\usepackage[framemethod=tikz]{mdframed}
\usepackage{multicol}
\usepackage{environ}
\usepackage{tikz}
\begin{document}
\title{
Dual-Functional FMCW Waveform for Terahertz Space Debris Detection and Inter-Satellite Communications
}
\author{
\IEEEauthorblockN{$\textrm{Zhepu Yin}$, $\textrm{Weijun Gao}$, and $\textrm{Chong Han}$}
\IEEEauthorblockA{Shanghai Jiao Tong University, Shanghai, China. E-mail:
\{zhepu.yin; gaoweijun; chong.han\}@sjtu.edu.cn\\
}
}

\markboth{}
\MakeLowercase
\maketitle
\begin{abstract}
\boldmath
Terahertz (THz) band communication, ranging from $0.1~\textrm{THz}$ to $10~\textrm{THz}$, is envisioned as a key enabling technology for next-generation networks and future applications such as inter-satellite communications and environmental sensing. The surging number of space debris in Low Earth Orbit poses a big threat to orbital infrastructure and the development of the space economy. In particular, despite the ability to detect and track large-scale space debris, millions of space debris with a radius within the range of $0.1-10~\textrm{cm}$ and velocity exceeding $1~\textrm{km}/\textrm{s}$ remains hard to detect with conventional ground-based radars and optical telescopes. In this study, a dual-functional frequency modulated continuous waveform (FMCW) operating in the THz band is adopted for space debris sensing and inter-satellite communications. Specifically, the radar cross section of space debris with various sizes in the THz band is analyzed to demonstrate the feasibility of THz space debris detection. A joint space debris detection and inter-satellite communications based on the FMCW waveform is derived. Then, the parameter estimation and demodulation algorithms are illustrated. Extensive simulations demonstrate that the proposed method can realize high-accuracy parameter estimation of hypervelocity space debris while achieving high reliability for inter-satellite communications.  
\end{abstract}

\maketitle

\section{Introduction}
Terahertz (THz) band communication, ranging from $0.1~\textrm{THz}$ to $10~\textrm{THz}$, is envisioned as a key enabling technology for next-generation networks and future applications such as inter-satellite communication. Thanks to their large continuous bandwidth at hundreds of gigahertz and small wavelength at several millimeters, THz band communication simultaneously possesses the ability of sensing and communication. Recently a real THz communication system is implemented to verify the viability of achieving THz high-speed data transfer~\cite{liu2024high}. These advancements motivate extensive applications based on THz communications, among which, inter-satellite communications~\cite{chen2019survey} is the new frontier. On one hand, THz inter-satellite communications are immune from water vapor absorption, which is considered as one of the key limitations of the THz terrestrial long-distance communications but not an issue in space. On the other hand, high directionality endows THz communications with strong resistance to malicious interception yet does not make it vulnerable to sunlight irradiance like its optical counterpart. 
Pioneering studies have verified the feasibility of THz inter-satellite communications in~\cite{li2021propagation,nie2021channel}.
Besides the advantages of the THz band in inter-satellite communications, space debris detection, which aims to detect and track the movement of objects in space, is one of the most potential applications for THz radar sensing. 

Since Sputnik, the first artificial satellite in space, was launched in 1957, the number of objects launched into space has increased steadily, surging from around 14,698 in 2015 to 34,595 by 2023~\cite{goel2016detection}.
Euroconsult forecasts an average of over 2,900 satellite launches annually, equivalent to 8 satellites per day and totaling a mass of 4 tons, in the next decade~\cite{carrasquilla2019debrisat,murray2019haystack}. 
Due to mission-related operations, accidents, or intentional creation like Anti-Satellite (ASAT) weapon attacks, a large amount of non-functional artificial objects, known as space debris, have been generated and drifted haphazardly in Earth orbit. 
As more and more space debris occurs, satellites are at increasing risk of collision with debris, leading to complete destruction or fatal damage.
Given the degrading space environment, the US Space Surveillance Network (SSN) has made efforts to detect and track space debris. However, small space debris with a radius between $0.1-10~\textrm{cm}$ is undetectable with existing ground-based radar sensing technology, and these small objects can pose a great threat to satellites due to their hypervelocity ($1~\textrm{km}/\textrm{s}$) and massive amount. Statistics show that there are around 129 million debris objects within the range of $0.1-10~\textrm{cm}$ and an aluminum sphere of $7~\textrm{km}/\textrm{s}$ can penetrate completely through an aluminum plate $4$ times its diameter. 
Due to the sub-millimeter-scale wavelengths, small-scale space debris with radius of several centimeters possesses higher radar cross section (RCS) in the THz band than in the lower-frequency bands.
To address the aforementioned challenges, in this paper, we develop a dual-functional FMCW waveform for THz space debris detection and inter-satellite communications, which simultaneously realize fast-speed data transfer among satellites and accurate distance and speed estimation for hypervelocity space debris. Specifically, a system model is first developed, where a THz line-of-sight (LoS) channel model for both functions is described. Then, the dual-functional FMCW waveform is designed and the receiver design for both communication and radar sensing purposes is elaborated. Finally, we perform extensive numerical evaluations to verify the communication and sensing performance of our dual-functional FMCW waveform, in terms of the bit-error rate (BER) and their sensing distance and speed root mean square error (RMSE).

The rest of the paper is organized as follows. In Sec.~\ref{sec:system}, a system model for joint space debris detection and inter-satellite communications based on dual-functional FMCW waveform is developed, where the channel model is developed in~\ref{sec:channel}, the dual-functional FMCW is designed in~\ref{sec:FMCW}, and the receiver design for communication and detection is elaborated in~\ref{sec:receiver}. 
Numerical results are described and analyzed in Sec.~\ref{sec:NR}. Finally, Sec.~\ref{sec:conc} concludes the paper. 

\section{System Model for Terahertz Space Debris Detection and Inter-Satellite Communications}~\label{sec:system}
In this section, we describe the system model for hypervelocity space debris detection based on the dual-functional FMCW waveform for joint space debris detection and inter-satellite communication. 
\begin{figure*}[t]
    \centering
    \includegraphics[width=1\linewidth]{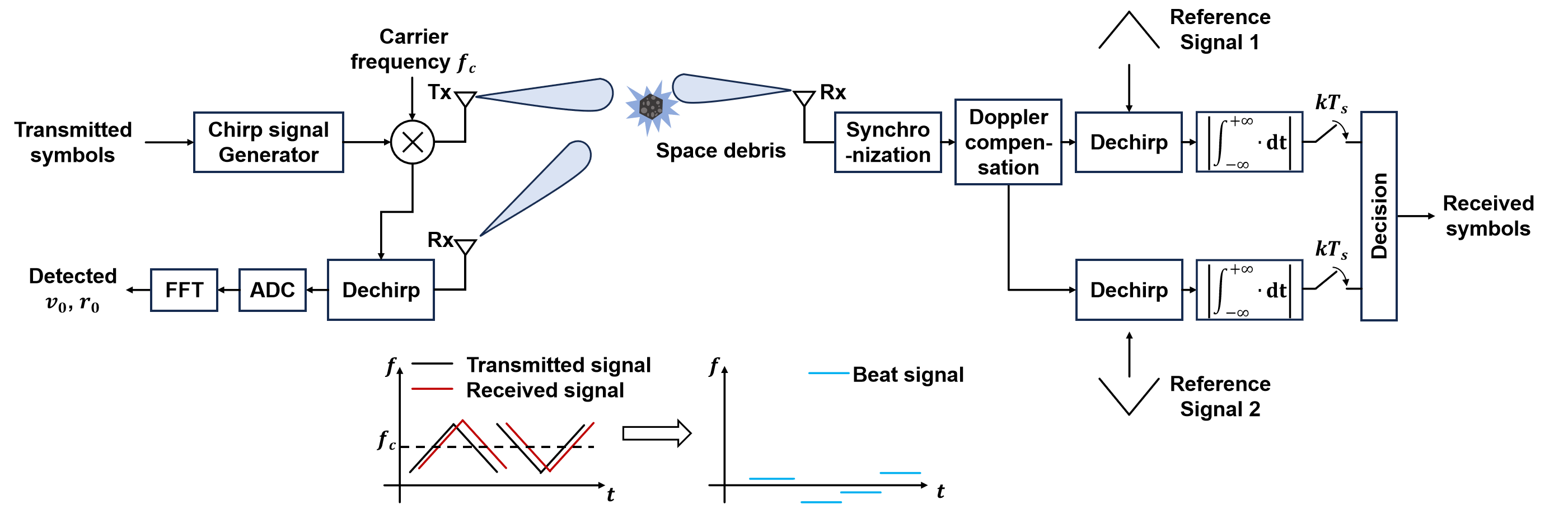}
    \caption{System model for THz joint space debris detection and inter-satellite communication.}
    \label{fig:system}
\end{figure*}
The monostatic integrated communication and radar sensing system block diagram is illustrated in Fig.~\ref{fig:system}, where the chirp signal generator on board the satellite transmits a THz FMCW signal carrying the encoded symbols. For the communication part, the receiver detects the transmitted signal and demodulates the binary frequency-modulated symbols for information transmission. For the sensing part, the space debris reflects the incoming THz signal, and the echo signal is captured by the sensing receiver colocated at the transmitter side. Then, the sensing receiver tries to estimate the distance and speed of space debris for further trajectory tracking and prediction. 
\subsection{Terahertz LoS Channel Model for Space Debris Detection and Inter-Satellite Links}~\label{sec:channel}
Due to the lack of scatters in the empty space, we assume that the THz channel model for space debris detection only possesses the LoS ray, where the communication receiver directly receives the signal through a LoS path, and the sensing receiver detects the signal which experiences single reflection by the space debris. Moreover, due to the lack of medium molecules such as water vapor or oxygen, this THz LoS channel assumes that the LoS ray only experiences free-space path loss (FSPL), and the molecular absorption effect is not considered~\cite{yang2024universal}. Therefore, the channel impulse response (CIR) of the THz communication channel $h_{ct}(t,\tau)$ is expressed as
\begin{equation}
\begin{aligned}
        \ h_{c}(t,\tau)&=\alpha_{LoS}\delta(\tau-\tau_{LoS}(t)),
\end{aligned}
\end{equation}
where $t$ denotes the time. $\delta(\cdot)$ represents the Dirac Delta function. $\tau_{LoS}(t)$ is the time-varying delay between the satellites along the same orbit and is calculated by $\tau_{LoS}(t)=\frac{R+vt}{c}$. $c$ is the speed of light, $R$ represents the distance between the satellites, and $v$ represents the relative speed. $\alpha_{LoS}$ stands for the FSPL for the THz LoS ray, which can be computed as
\begin{equation}
\begin{aligned}
        \alpha_{LoS}=\frac{c}{4\pi f_c R},
\end{aligned}
\end{equation}
where $f_c$ represents the carrier frequency.
For the sensing part, the THz CIR of the sensing channel is described as
\begin{equation}
    h_{s}(t,\tau)=\alpha_{s}\delta(\tau-\tau_{s}(t)),
\end{equation}
where the time-varying delay for the two-way propagation shown in Fig.~\ref{fig:system} can be calculated by $\tau_s(t)=\frac{2(r_0+v_0 t)}{c}$. $r_0$ is the distance between the transmitter and the detected space debris, and $v_0$ denotes the speed of the space debris.
Similar to the THz communication channel model, the FSPL for the sensing signal is calculated as
\begin{equation}
\begin{aligned}
        \alpha_{s}=\frac{c}{4\pi f_c r_0}\cdot \sqrt{\frac{\sigma}{4\pi r_0^2}},
\end{aligned}
\end{equation}
where $\sigma$ represents the RCS of the space debris in the THz band. According to the NASA size estimation model (SEM)~\cite{murray2019haystack}, the RCS is strongly related to the size of the space debris.
For space debris whose radius is much smaller than the THz wavelengths, RCS can be calculated as the Rayleigh scattering model, while for the case that the radius is longer than the wavelengths, the geometric scattering model is adopted. For the remaining regions between the two regions, the RCS curve is polynomial fitted~\cite{murray2019haystack} using $23$ measurement results provided by NASA. Therefore, the RCS model for these cases is given by
\begin{equation}
    \ \hat{\sigma}=
    \begin{cases}
        x^2/(4\pi),& x>2.523,\\
        9\pi^2x^6/4,&  x<0.1876,\\
        \min(a_0x+b_0, c_0x+d_0, e_0x+f_0),& \textrm{otherwise},
    \end{cases}
    \label{eq:RCS}
\end{equation}
where $\hat{\sigma}$ is the normalized RCS, i.e., $\hat{\sigma}=\frac{\sigma}{\lambda^2}$. $x$ denotes the normalized radius of space debris, i.e., $x=\frac{r}{\lambda}$. And $\lambda=\frac{c}{f_c}$ represents the wavelength. $\min(\cdot)$ returns the minimum value among the three input values. The fitting coefficients for $x$ in~\eqref{eq:RCS} are given by 
\begin{equation}
\begin{aligned}
    a_0&=11.0346, b_0=-11.2369,\\
    c_0&=4.9409, d_0=-4.8158,\\
    e_0&=157.2653, f_0=-45.6594.\\
\end{aligned}
\end{equation}

\begin{figure}[ht]
\centering
\includegraphics[width=0.8\linewidth]{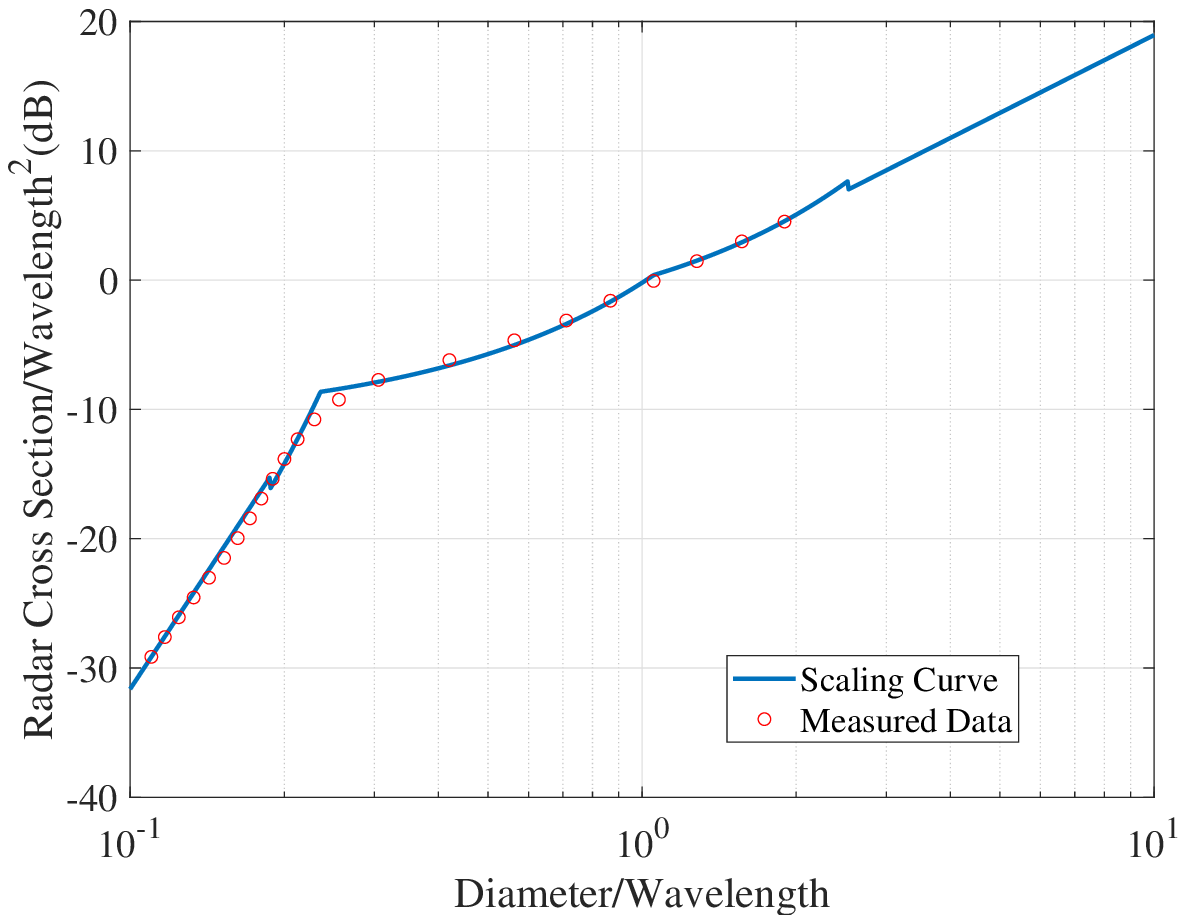}
\captionsetup{font={footnotesize}}
\caption{RCS static range measurements and its polynomial fit as a function of scaling parameters $x=\frac{r}{\lambda}$ and $z=\frac{\sigma}{\lambda^2}$.}
\label{fig:RCSCurve}
\end{figure}
The scaling RCS of space debris is shown in Fig.~\ref{fig:RCSCurve}. 
We observe that as the diameter of space debris increases and the wavelength decreases, the RCS of space debris grows accordingly, which implies that compared with micro-wave or millimeter-wave frequency bands, the THz band possesses a better detection ability of space debris due to its smaller wavelength. Furthermore, an operative frequency of $340~\textrm{GHz}$ is not only widely used for radar sensing according to~\cite{cerutti2017preliminary,yang2017three}, but also guarantees that small space debris larger than $2.2~\textrm{mm}$ is in the optical regime. This result verifies the feasibility of using THz wave for space debris detection.
 
\subsection{Dual-Functional FMCW Waveform Design}~\label{sec:FMCW}
We develop a waveform model for the dual-functional FMCW waveform. The transmitted signal is expressed as
\begin{equation}
\begin{aligned}
        \ s_t(t)&=\exp\left[j2\pi \left(f_c t+\frac{1}{2}\mu t^2\right)\right], 0\leq t\leq T_c,
\end{aligned}
\end{equation}
where $\mu=\frac{B_c}{T_c}$ denotes the chirp rate, which is determined by the chirp duration $T_c$ and the bandwidth of the THz FMCW signal $B_c$. By neglecting the additive white Gaussian noise (AWGN), free-space path loss, and reflection loss, the echo signal received by the FMCW receiver is a delayed version of the transmitted $s_t(t)$, which is therefore given by
\begin{equation}
\begin{aligned}
        s_r(t)=\exp\left[j2\pi(f_c (t-\tau(t))+\frac{1}{2}\mu (t-\tau(t))^2\right],
\end{aligned}
\end{equation}
where $\tau(t)$ is the time-varying two-way time delay, $r_0$ is the initial distance of the target at $t=0$, and $v_0$ is the constant speed. At the sensing receiver, a dechirp process is performed, and the resulting beat signal for distance and speed estimation is expressed as
\begin{equation}
\begin{aligned}
        &s_{beat}(t)=s_t(t)\cdot s^{*}_r(t) \\
        &=\exp\left[j2\pi \left(\mu \tau t+f_c\tau-\frac{1}{2}\mu\tau^2\right)\right]\\
        &=\exp\Bigg\{j2\pi \Bigg[\frac{2\mu v_0}{c}(1-\frac{v_0}{c})t^2+\frac{2}{c}\left(f_cv_0+\mu r_0-\frac{2\mu r_0 v_0}{c}\right)t\\
        &+\frac{2f_cr_0}{c}-\frac{2\mu r_0^2}{c^2} \Bigg] \Bigg\}.
\end{aligned}
\label{eq:beat}
\end{equation}

In conventional analysis with a low speed of detected targets, $s_{beat}(t)$ is considered as a single-tone sinusoid signal by omitting the second-order term and the $-\frac{2\mu r_0v_0}{c}$ term in the first-order term in~\eqref{eq:beat}. Through this approximation, the distance and speed estimation problem of the slow target is converted into the problem of frequency estimation, which, in this case, is the coefficient of the first-order term, i.e., $f_cv_0+\mu r_0$. 
However, in the scenario of space debris detection with a hypervelocity more than $1000~\textrm{km}/\textrm{s}$, the detection precision is limited as the phase of the beat signal is a quadratic function, whose coefficient of the quadratic term is determined by the speed of the target and cannot be neglected. 

To address this issue, we modulate the FMCW in a triangular shape, which contains an up-chirp and a down-chirp in a symbol duration. Thus, the distance and speed of the target can be estimated using a triangular chirp and the quadratic term in phase can be neglected since the frequency change is rather small. The beat frequency of the up-chirp and down-chirp can be expressed as
\begin{align}
        \ f_{beat, up}&=\frac{2}{c}\left(f_cv_0+\mu r_0-\frac{2\mu r_0v_0}{c}\right),
        \label{eq:beat,up}
        \\
        \ f_{beat, down}&=\frac{2}{c}\left(f_cv_0-\mu r_0+\frac{2\mu r_0v_0}{c}\right),
\label{eq:beat,down}
\end{align}

Thus, by combing  ~\eqref{eq:beat,up} and ~\eqref{eq:beat,down}, the distance and speed can be computed as
\begin{align}
        v_0&=\frac{c}{4f_c}(f_{beat,up}+f_{beat,down}),
        \label{eq:v0}
        \\
        r_0&=\frac{c}{4\mu(1-2v_0/c)}(f_{beat,up}-f_{beat,down}),
        \label{eq:r0}
\end{align}

To enable FMCW with an information-conveying function, the waveform is modulated in a manner similar to conventional pulse-based waveform, where the polarization of the FMCW determines the information bit. In particular, if we want to transmit a bit ``0", then we can send an up-down-shaped signal. While if we want to transmit a bit ``1", we just need to change the orientation, i.e., transmit a down-up-shaped signal. Note that despite the proposed communication scheme being less efficient than conventional modulations, it is better than the sensing-only scheme with FWCM signals.

\subsection{Receiver for Communication and Sensing}\label{sec:receiver}
As shown in Fig.~\ref{fig:system}, the sensing receiver is co-located with the transmitter. After dechirping, a Fast Fourier Transform (FFT) operation is performed on the beat signal, and the signal frequency is estimated by finding the peak in the periodogram.  Moreover, a $N$-point FFT operation on the single-tone signal is equivalent to filtering the signal with a filter bank consisting of a $N$ narrowband filter. With the power of signal increasing by $N^2$ times while that of noise increases by $N$ times, the signal-to-noise ratio (SNR) increases by around $N$ fold. To further improve the resolution in the frequency domain, we use zero padding by a factor of two. Finally, the estimated frequency is used to calculate the distance and speed of the target as described in~\eqref{eq:v0} and~\eqref{eq:r0}.

For the communication receiver, due to the long distance of ISL and fast relative movement between transmitter and receiver, the delay and Doppler frequency shift should be compensated first before demodulation. Then, the synchronized signal is dechirped with the reference waveform. The output is integrated over a symbol duration as 
\begin{equation}
    k_{up/down}=\left|\int_{0}^{T_C}s_{beat}(t)dt\right|,
    \label{eq:integrate}
\end{equation}
and the received bit $b_r$ is determined by comparing the output values $k_{up}$ and $k_{down}$.
Based on this system, we can simultaneously maintain the ability to accurately detect hypervelocity and small-scale space debris.

\section{Numerical Results}\label{sec:NR}
In this section, we conduct numerical evaluations on the proposed FMCW waveform for THz space debris detection and inter-satellite communication.  Specifically, the sensing accuracy versus SNR is evaluated and analyzed for the different nominal distances and speeds of space debris. To investigate the performance of inter-satellite communications, the BER versus SNR is computed. Unless specified, the simulation parameters are listed in Table.~\ref{Table}. Note that the frequency is set as $340~\textrm{GHz}$, which is widely used for radar sensing according to~\cite{cerutti2017preliminary,yang2017three}.
    \begin{table}[t] 
        \caption{Simulation Parameters} 
        \label{Table}
        \centering
        \begin{tabular}{p{1.5cm}p{3cm}p{1.7cm}p{0.8cm}} 
            \hline  
            \hline  
            \textbf{Notation} & \textbf{Parameter Definition} & \textbf{Value} & \textbf{Unit} \\ 
            \hline 
            $h$ & Orbit altitude & 600-1000 & km\\
            $R$ & Communication distance & 200-700 & km\\
            $v$ & Relative speed & 7 & km/s\\
            $d$ & Diameter of space debris & 0.1-10 & cm\\
            $r_{max}$ & Maximum distance & 500/2000 & m\\
            $v_{max}$ & Maximum speed & $\leq$ 15 & km/s\\
            $r_{res}$ & Distance resolution & 0.1-10 & cm\\
            $f_c$ & Carrier frequency & $340$ & GHz\\
            $T_c$ & Chirp duration & $18.35$ & $\mu s$\\
            $B_c$ & Bandwidth & $1.5$ & GHz\\
            \hline
            \hline  
        \end{tabular}  
    \end{table} 

\subsection{Performance Evaluation for Space Debris Detection}
\begin{figure}[ht]
\centering
\includegraphics[width=0.8\linewidth]{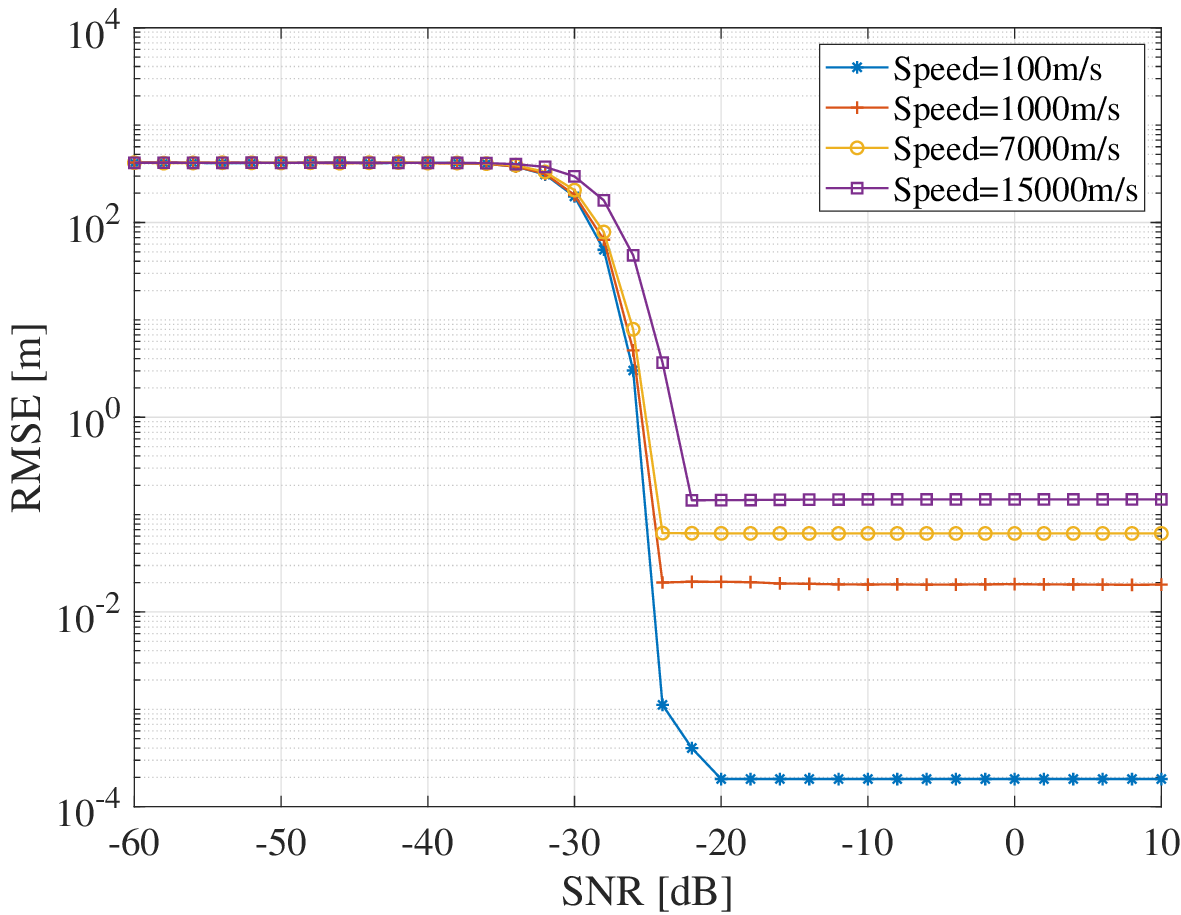}
\captionsetup{font={footnotesize}}
\caption{Comparison of distance estimation accuracy over different target speeds.}
\label{fig:RangeVersusVelocity}
\end{figure}

\begin{figure}[ht]
\centering
\includegraphics[width=0.8\linewidth]{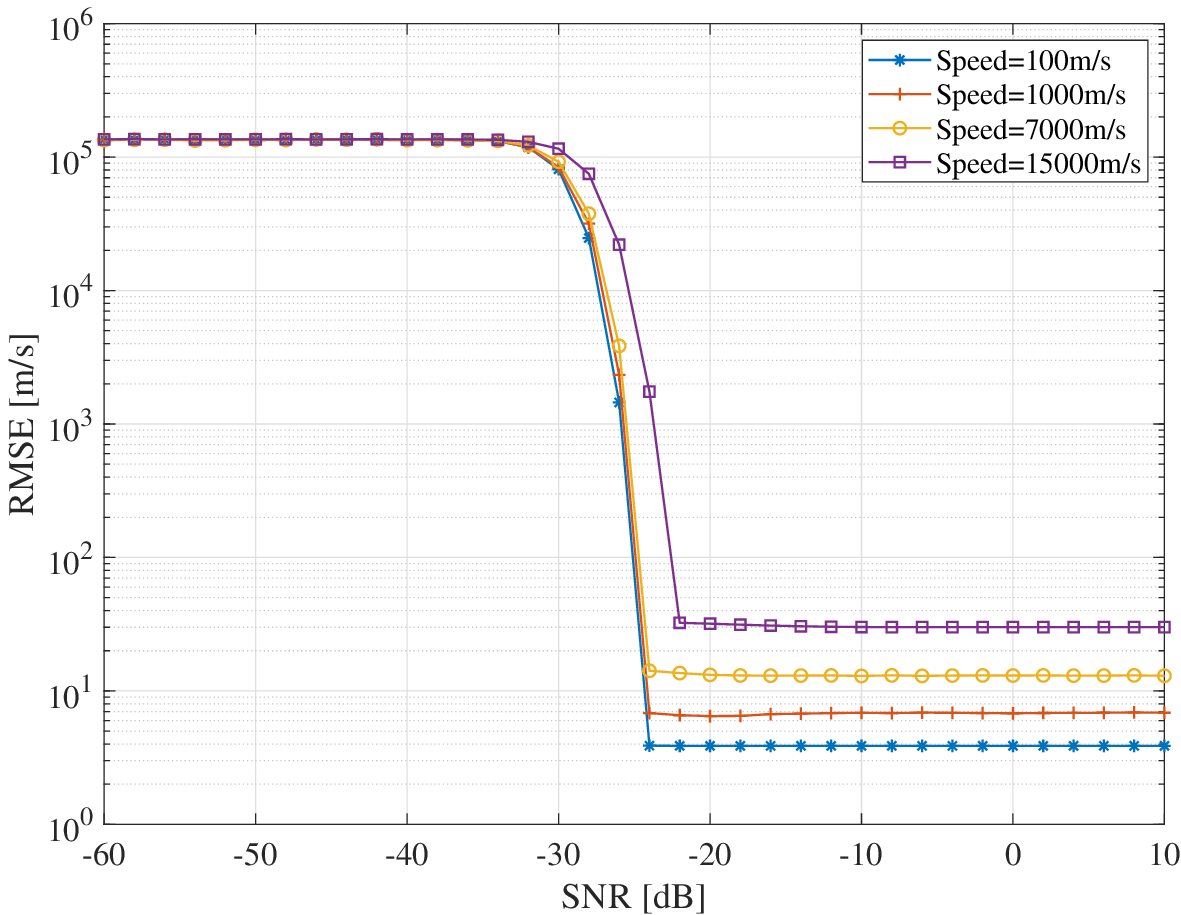}
\captionsetup{font={footnotesize}}
\caption{Comparison of speed estimation accuracy over different target speeds.}
\label{fig:VelocityVersusVelocity}
\end{figure}

In Fig.~\ref{fig:RangeVersusVelocity} and Fig.~\ref{fig:VelocityVersusVelocity}, we evaluate the RMSE of distance and speed estimation across different SNR and target speeds with $r_{max}=500~\textrm{m}$ and target distance fixed at $300~\textrm{m}$. We observe that as the target speed increases, the estimation accuracy decreases. For distance estimation, it is indicated that the RMSE achieves below $0.15~\textrm{m}$ at  $v=15~\textrm{km}/\textrm{s}$. When $v\approx 7~\textrm{km}/\textrm{s}$, the distance estimation accuracy can achieve centimeter-level. 
For speed estimation, the maximum RMSE is around $30~\textrm{m}/\textrm{s}$ when $v_0=15~\textrm{km}/{s}$ while is average is below  $15~\textrm{m}/\textrm{s}$ for all considered speed.
As for anti-noise capability, the SNR threshold in the simulations is at around $-24~\textrm{dB}$, which is $5/6/7~\textrm{dB}$ lower than that of OFDM/FMCW/OTFS based on maximum likelihood estimation algorithm, respectively. This indicates that the triangular modulated FMCW and corresponding estimation algorithm are more robust to additive noise than other reference waveforms. 

\begin{figure}[ht]
\centering
\includegraphics[width=0.8\linewidth]{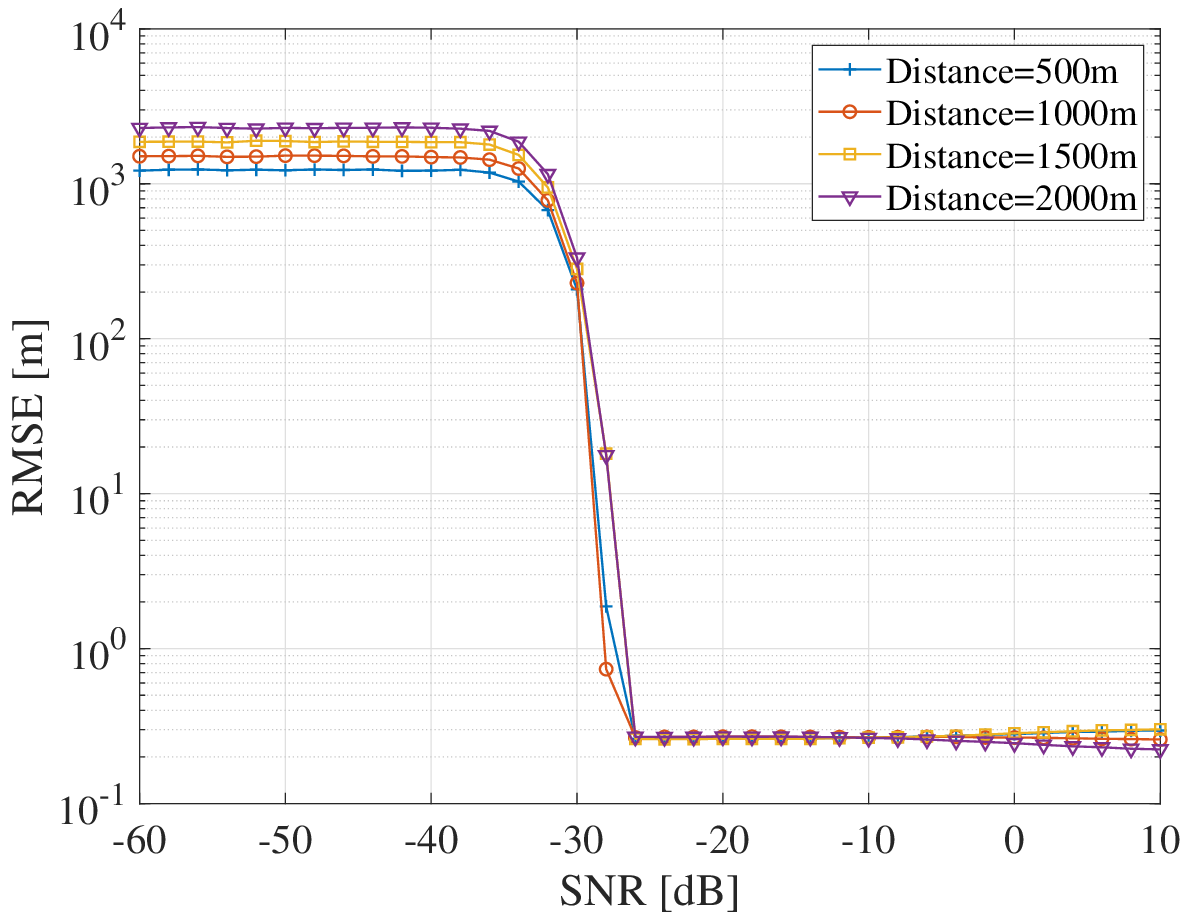}
\captionsetup{font={footnotesize}}
\caption{Comparison of distance estimation accuracy over different target distances.}
\label{fig:RangeVersusRange}
\end{figure}

\begin{figure}[ht]
\centering
\includegraphics[width=0.8\linewidth]{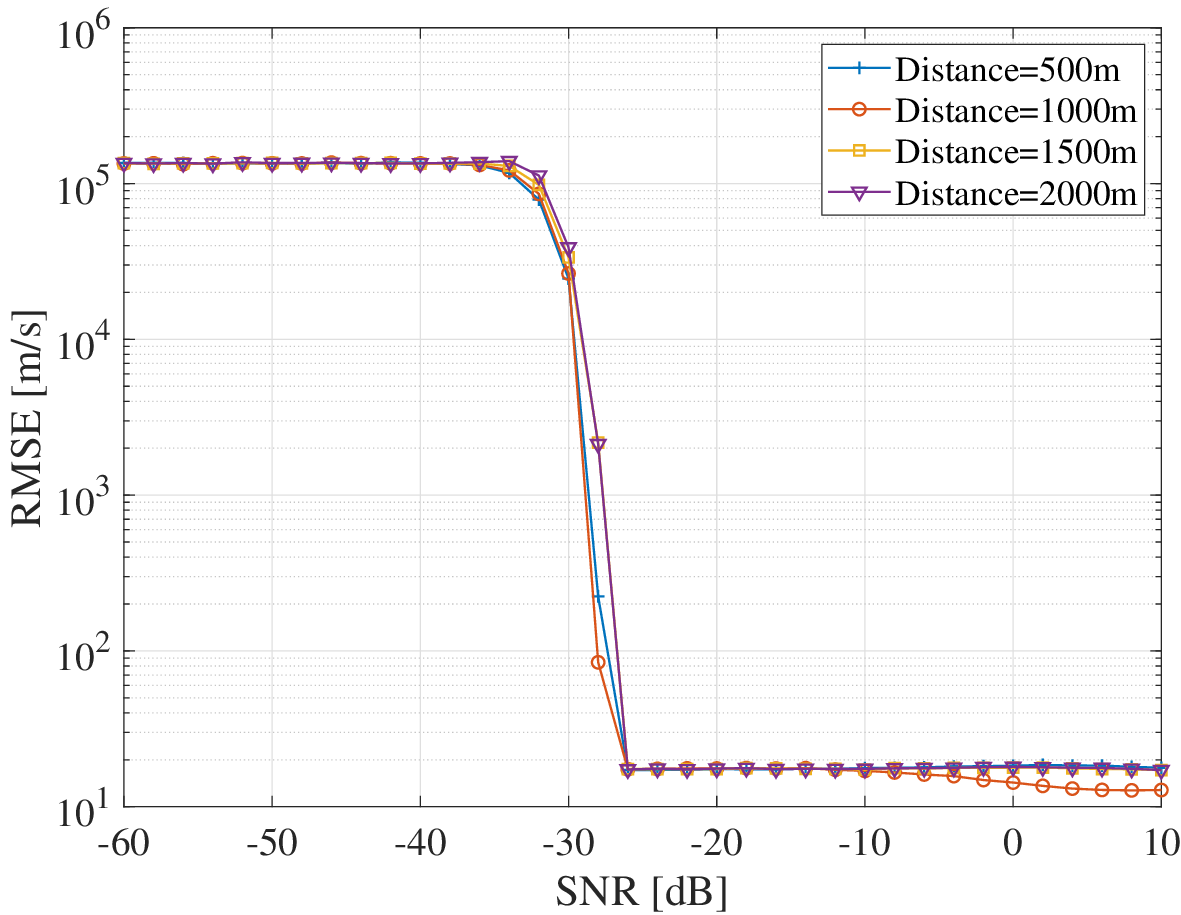}
\captionsetup{font={footnotesize}}
\caption{Comparison of speed estimation accuracy over different target distances. }
\label{fig:VelocityVersusRange}
\end{figure}

Moreover, we evaluate the influence of target distance on the RMSE of distance and speed estimation with $r_{max}=2000~\textrm{m}$ and target speed fixed at $7~\textrm{km}/\textrm{s}$. In Fig.~\ref{fig:RangeVersusRange} and Fig.~\ref{fig:VelocityVersusRange}, we observe that the target distance has little effect on sensing accuracy, with distance estimation accuracy approaching decimeter-level and speed estimation RMSE below $20~\textrm{m}/\textrm{s}$.

\begin{figure}[ht]
\centering
\includegraphics[width=0.75\linewidth]{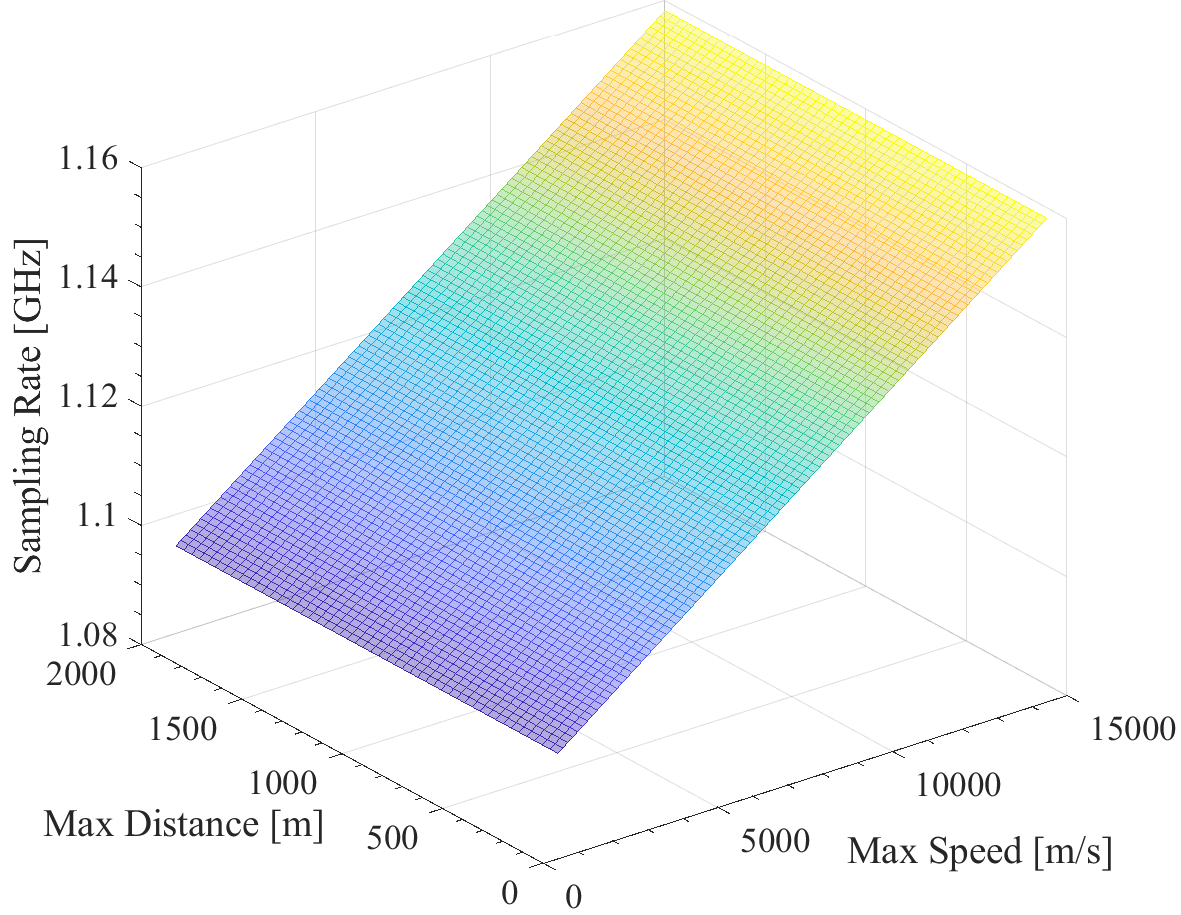}
\captionsetup{font={footnotesize}}
\caption{Sampling Rate versus maximum detectable distance and speed.}
\label{fig:SamplingRate1}
\end{figure}

\begin{figure}[ht]
\centering
\includegraphics[width=0.75\linewidth]{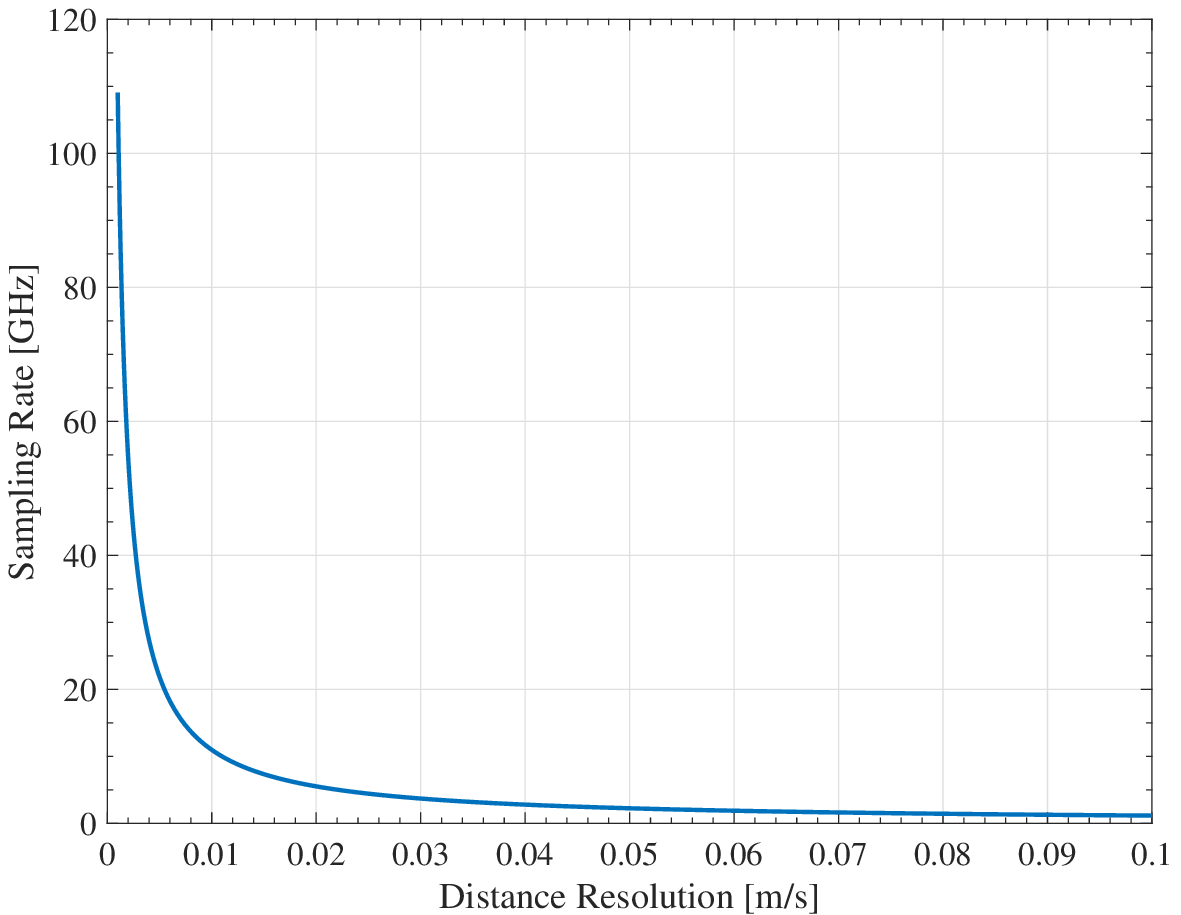}
\captionsetup{font={footnotesize}}
\caption{Sampling Rate versus distance resolution.}
\label{fig:SamplingRate2}
\end{figure}

As shown in Fig.~\ref{fig:SamplingRate1}, the sampling rate increases linearly with the maximum detectable speed while it is irrelevant with unambiguous distance. In particular, when distance resolution is fixed at $0.1~\textrm{m}$, the sampling rate is below $1.2~\textrm{GHz}$ even with the maximum detectable speed at $15~\textrm{km/s}$. This property brought by dechirp operation indicates that the receiver can be built with off-the-shelf components instead of ADCs operating at hundreds of Giga-samples-per-second (GSaps)\cite{akyildiz2022terahertz}, thus cutting down the total cost. In Fig.~\ref{fig:SamplingRate2}, results demonstrate that sampling rate surges explosively if distance resolution is below $0.01~\textrm{m}$, achieving $110$~GSaps at peak. In practical hardware implementation, a higher sampling rate will significantly increase system complexity like heavier computation loads, larger storage requirements, larger power consumption, etc. In consideration of the system's real-time performance required by the hypervelocity of space debris and system cost, the trade-off between system complexity and distance resolution is crucial.

\begin{figure}[ht]
    \centering
    \includegraphics[width=0.75\linewidth]{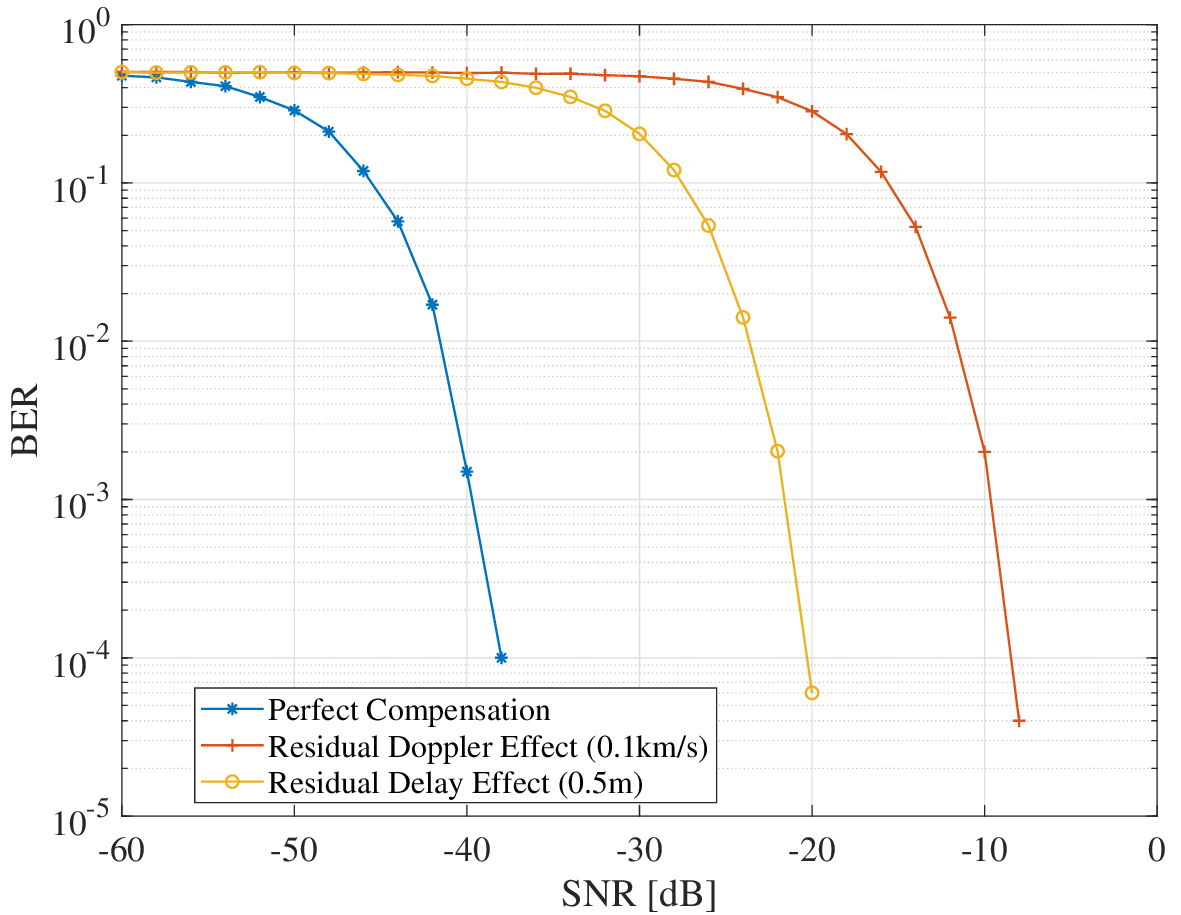}
    \captionsetup{font={footnotesize}}
    \caption{Comparison of BER under different compensation Conditions.}
    \label{fig:BER}
\end{figure}

\subsection{Performance Evaluation for Inter-Satellite Communications}
We compare BER performance under different compensation conditions, i.e. perfect delay and Doppler compensation, perfect delay compensation with $0.227~\textrm{MHz}$ residual Doppler shift, and perfect Doppler compensation with $1.67~\textrm{ns}$ residual delay. Here, $0.227~\textrm{MHz}$ residual Doppler shift corresponds to $0.1~\textrm{km}/\textrm{s}$ in speed while $1.67~\textrm{ns}$ residual delay corresponds to $0.5~\textrm{m}$ in distance. As shown in Fig.~\ref{fig:BER}, the simulation results indicate that under perfect Doppler and delay compensation, high demodulation accuracy can be obtained with $\textrm{SNR}=-38~\textrm{dB}$ at the BER of $10^{-4}$, which means the required power consumption is low for reliable communications in ISLs. This exactly meets the strict demand of CubeSat, whose transmitting power is typically between $15~\textrm{dBm}$ and $7~\textrm{dB}$~\cite{hodges2017deployable}. However, provided no perfect compensation, the demodulation degrades drastically. In particular, when $0.227~\textrm{MHz}$ Doppler shift exists, $37.5~\textrm{dB}$ more power should be transmitted to maintain BER at the $10^{-4}$ level. The degradation is more serious when the delay is not compensated, a $1.67~\textrm{ns}$ residual delay needs $20~\textrm{dB}$ extra power to achieve the $10^{-4}$ level BER.
Therefore, precise synchronization in the time-frequency domain is vital for FMCW-based communication systems.

\begin{figure}[ht]
\centering
\subfigure[]{
\label{fig:datarate1} 
\includegraphics[width=0.75\textwidth/2]{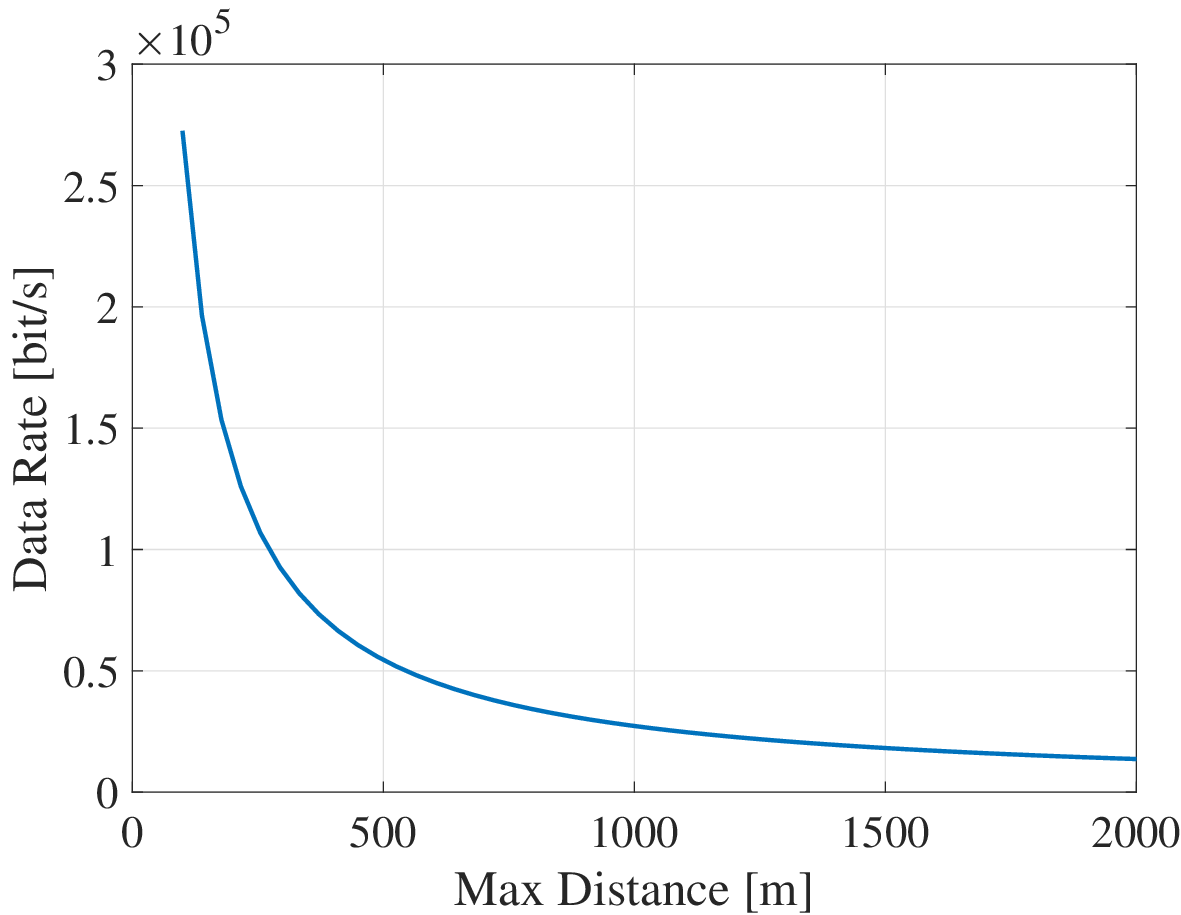}}
\subfigure[]{
\label{fig:datarate2}
\includegraphics[width=0.75\textwidth/2]{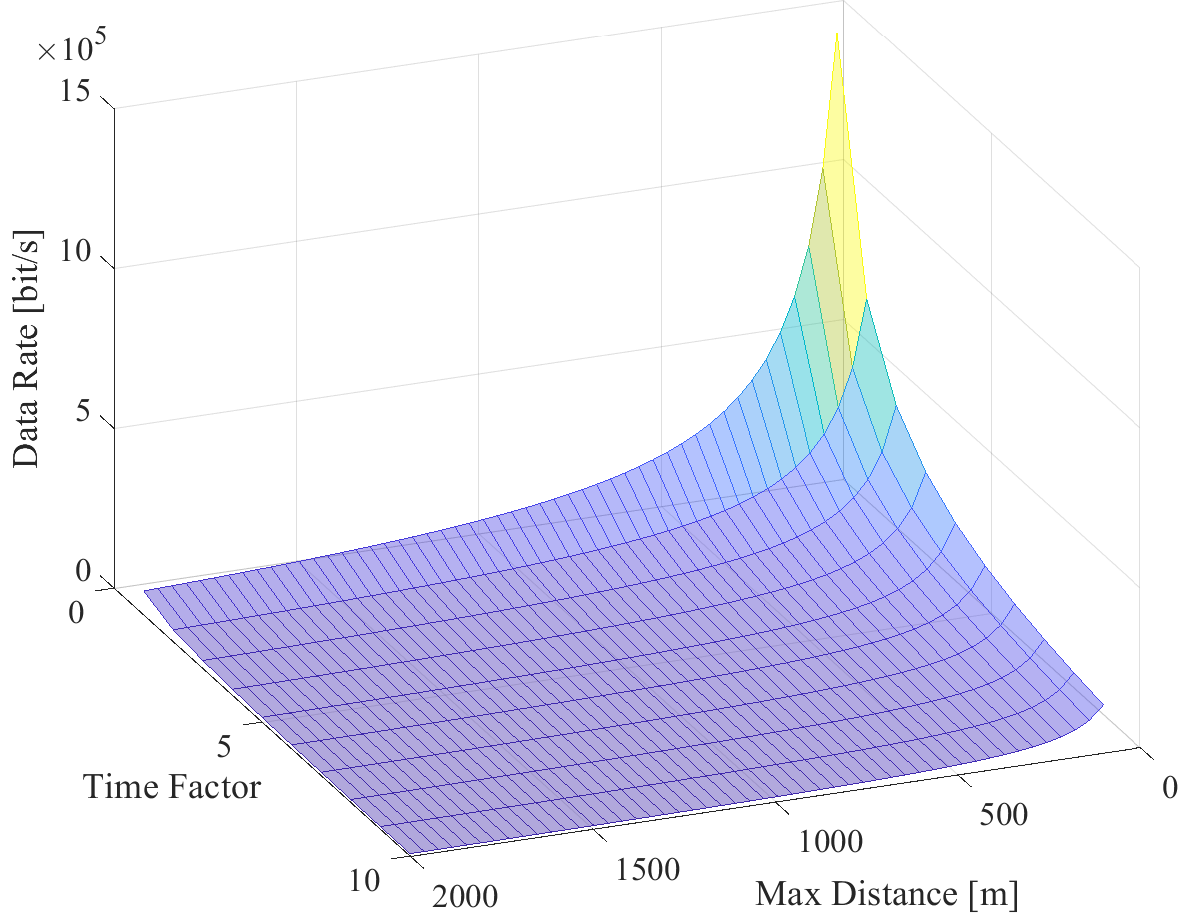}}
\captionsetup{font={footnotesize}}
\caption{Transmitted bits per second (a) under different unambiguous distance settings with time factor 5.5; (b) versus different unambiguous distance and time factor settings.}
\label{fig:datarate} 
\end{figure}

We further simulate the data rate of the THz inter-satellite communication system, which is defined as the number of bits transmitted over $1$ second. As shown in Fig.~\ref{fig:datarate1}, a data rate at $50~\textrm{kbps}$ can be achieved if the unambiguous distance is $500~\textrm{m}$. Moreover, the data rate decreases with the increase of unambiguous distance, which shows a trade-off between communication and sensing performance. The relationship between $T_c$ and $r_\textrm{max}$ is expressed as $T_c=a\cdot2r_{max}/c$, where $a$ is the protection time factor to make sure the chirp duration is large enough compared with the maximum round-trip time delay and is set as 5.5 experientially in automobile radar. Hence, to detect farther space debris means longer chirp duration, thus lowering the data rate. Such data rata can only satisfy applications like telemetry data transmission, low-resolution image transmission, and control links. Although decreasing the time factor can shorten chirp duration, this may be at the expense of degrading sensing performance, which is a trade-off for joint space debris detection and inter-satellite communications. As shown in Fig.~\ref{fig:datarate2}, when $a$ is tuned to 1, the data rate can achieve $1.5~\textrm{Mbps}$. In view of broadband applications like high-definition video streaming or large file transfers, more modifications should be made to FMCW to improve its communication capability.
\section{Conclusion}~\label{sec:conc}
In this paper, we design a dual-functional FMCW waveform operating in the THz band for space debris sensing and inter-satellite communication. Specifically, the radar cross section of space debris with various sizes in the THz band is analyzed to demonstrate the feasibility of THz space debris detection. A theoretical system model based on the FMCW waveform for joint space debris detection and inter-satellite communications is derived. Then, the parameter estimation and demodulation algorithms are illustrated. Extensive simulations demonstrate that the proposed method can realize  decimeter level distance estimation accuracy and ten-meter-per-second level speed estimation of ultra-fast space debris. In addition, the system can achieve high reliability for ISLs since only $-38$~dB SNR is needed to reduce BER to $10^{-4}$ level.  

\bibliographystyle{ieeetr}
\bibliography{main}

\begin{thebibliography}{10}

\bibitem{liu2024high}
K.~Liu, Y.~Feng, C.~Han, B.~Chang, Z.~Chen, Z.~Xu, L.~Li, B.~Zhang, Y.~Wang,
  and Q.~Xu, ``High-speed 0.22 thz communication system with 84 gbps for
  real-time uncompressed 8k video transmission of live events,'' {\em Nature
  Communications}, vol.~15, no.~1, p.~8037, 2024.

\bibitem{chen2019survey}
Z.~Chen, X.~Ma, B.~Zhang, Y.~Zhang, Z.~Niu, N.~Kuang, W.~Chen, L.~Li, and
  S.~Li, ``A survey on terahertz communications,'' {\em China Communications},
  vol.~16, no.~2, pp.~1--35, 2019.

\bibitem{li2021propagation}
Y.~Li and Y.~Chen, ``Propagation modeling and analysis for terahertz
  inter-satellite communications using fdtd methods,'' in {\em 2021 IEEE
  International Conference on Communications Workshops (ICC Workshops)},
  pp.~1--6, IEEE, 2021.

\bibitem{nie2021channel}
S.~Nie and I.~F. Akyildiz, ``Channel modeling and analysis of
  inter-small-satellite links in terahertz band space networks,'' {\em IEEE
  Transactions on Communications}, vol.~69, no.~12, pp.~8585--8599, 2021.

\bibitem{goel2016detection}
A.~Goel, {\em Detection and characterization of meteoroid and orbital debris
  impacts in space}.
\newblock Stanford University, 2016.

\bibitem{carrasquilla2019debrisat}
R.~E. Carrasquilla and F.-C. Norman, ``Debrisat: Generating a dataset to
  improve space debris models from a laboratory hypervelocity experiment,''
  {\em Europe}, vol.~1, no.~361, pp.~5--638, 2019.

\bibitem{murray2019haystack}
J.~Murray, C.~Blackwell, J.~Gaynor, and T.~Kennedy, ``Haystack ultra-wideband
  satellite imaging radar measurements of the orbital debris environment:
  2014-2017,'' tech. rep., 2019.

\bibitem{yang2024universal}
Z.~Yang, W.~Gao, and C.~Han, ``A universal attenuation model of terahertz wave
  in space-air-ground channel medium,'' {\em IEEE Open Journal of the
  Communications Society}, 2024.

\bibitem{cerutti2017preliminary}
D.~Cerutti-Maori, J.~Rosebrock, I.~Maouloud, L.~Leushacke, and H.~Krag,
  ``Preliminary concept of a space-based radar for detecting mm-size space
  debris,'' in {\em Proc. of European Conf. on Space Debris}, 2017.

\bibitem{yang2017three}
X.~Yang, Y.~Pi, T.~Liu, and H.~Wang, ``Three-dimensional imaging of space
  debris with space-based terahertz radar,'' {\em IEEE Sensors Journal},
  vol.~18, no.~3, pp.~1063--1072, 2017.

\bibitem{akyildiz2022terahertz}
I.~F. Akyildiz, C.~Han, Z.~Hu, S.~Nie, and J.~M. Jornet, ``Terahertz band
  communication: An old problem revisited and research directions for the next
  decade,'' {\em IEEE Transactions on Communications}, vol.~70, no.~6,
  pp.~4250--4285, 2022.

\bibitem{hodges2017deployable}
R.~E. Hodges, N.~Chahat, D.~J. Hoppe, and J.~D. Vacchione, ``A deployable
  high-gain antenna bound for mars: Developing a new folded-panel reflectarray
  for the first cubesat mission to mars,'' {\em IEEE Antennas and Propagation
  Magazine}, vol.~59, no.~2, pp.~39--49, 2017.

\end{thebibliography}
\end{document}